\documentclass{article}

\PassOptionsToPackage{numbers}{natbib}
\usepackage[final]{nips_2016}

\usepackage[utf8]{inputenc} %
\usepackage[T1]{fontenc}    %
\usepackage{hyperref}       %
\usepackage{url}            %
\usepackage{booktabs}       %
\usepackage{amsfonts}       %
\usepackage{nicefrac}       %
\usepackage{microtype}      %

\usepackage{amsmath}
\usepackage{amssymb}
\usepackage{mathtools}
\DeclarePairedDelimiter\ceil{\lceil}{\rceil}

\usepackage{algorithmic}
\usepackage{array}
\usepackage{subfig}
\usepackage{todonotes}
\usepackage{xargs}

\newcommandx{\todofor}[3][1=]{}  %

\newcommandx{\todobonawitz}[2][1=]{\todofor[#1]{bonawitz}{#2}}
\newcommandx{\todovlivan}[2][1=]{\todofor[#1]{vlivan}{#2}}
\newcommandx{\todobenkreuter}[2][1=]{\todofor[#1]{benkrueter}{#2}}
\newcommandx{\todomcmahan}[2][1=]{\todofor[#1]{mcmahan}{#2}}
\newcommandx{\tododramage}[2][1=]{\todofor[#1]{dramage}{#2}}
\newcommandx{\todoasegal}[2][1=]{\todofor[#1]{asegal}{#2}}
\newcommandx{\todosarvar}[2][1=]{\todofor[#1]{sarvar}{#2}}
\newcommandx{\todokarn}[2][1=]{\todofor[#1]{karn}{#2}}
\newcommandx{\todomarcedone}[2][1=]{\todofor[#1]{marcedone}{#2}}

\newcommand{\reals}{{\mathbb R}}

\newcommand{\norm}{\ell}

\newcommand{\Update}{\delta}
\newcommand{\Database}{D}
\newcommand{\Parameters}{\Theta}
\newcommand{\Users}{{\cal U}}
\newcommand{\Batch}{B}
\newcommand{\Server}{S}
\newcommand{\Loss}{{\cal L}}  %

\newcommand{\numUsers}{n}
\newcommand{\numElements}{k}
\newcommand{\threshold}{t}
\newcommand{\maxElement}{R}

\DeclareMathOperator{\Agree}{\textsc{Agree}}
\DeclareMathOperator{\PRG}{\textsc{PRG}}

\usepackage{xcolor}

\newcommand{\threatmodel}[1]{T#1}

\newcommand{\user}{{u}}
\newcommand{\useri}{{u}}
\newcommand{\userj}{{v}}

\title{Practical Secure Aggregation for \\Federated Learning on User-Held Data}

\newcommand\Mark[1]{\textsuperscript#1}
\newcommand\MarkGoogle{\Mark{*}}
\newcommand\MarkCornell{\Mark{\textdagger}}

\author{
  Keith Bonawitz\MarkGoogle, Vladimir Ivanov\MarkGoogle, Ben Kreuter\MarkGoogle, Antonio Marcedone\MarkCornell\MarkGoogle, \\
  \textbf{H. Brendan McMahan\MarkGoogle, Sarvar Patel\MarkGoogle, Daniel Ramage\MarkGoogle, Aaron Segal\MarkGoogle, and Karn Seth\MarkGoogle} \\
  \MarkGoogle\texttt{\{bonawitz,vlivan,benkreuter,mcmahan,sarvar,dramage,asegal,karn\}@google.com} \\
  Google, Mountain View, California 94043\\
  \MarkCornell\texttt{marcedone@cs.cornell.edu} \\
  Cornell University, Ithaca, New York 14853
}

\begin{document}

\maketitle

\newcommand{\shortsection}[1]{\vspace{-0.1in}\section{#1}%
\vspace{-0.15in}}
\newcommand{\shortparagraph}[1]{\paragraph{#1}%
\vspace{-0.05in}}

\vspace{-0.1in}
\shortsection{Introduction}
{\em Secure Aggregation} is a class of Secure Multi-Party Computation algorithms
wherein a group of mutually distrustful parties $u \in \Users$
each hold a private value $x_u$ and collaborate to compute an aggregate
value, such as the sum $\sum_{u \in \Users} x_u$, without revealing to
one another any information about their private value except what is
learnable from the aggregate value itself.
In this work, we
consider training a deep neural network in the \emph{Federated Learning} model, using
distributed gradient descent across user-held training data on mobile devices,
using Secure Aggregation to protect the privacy of each user's model gradient.
We identify a combination of efficiency and robustness requirements which,
to the best of our knowledge, are unmet by existing algorithms in the literature.
We proceed to design a novel, communication-efficient Secure Aggregation protocol for
high-dimensional data that tolerates up to $\nicefrac{1}{3}$ of users failing to
complete the protocol.  For 16-bit input values, our protocol offers
$1.73\times$ communication expansion for $2^{10}$ users and
$2^{20}$-dimensional vectors, and $1.98\times$ expansion for $2^{14}$
users and $2^{24}$-dimensional vectors.

\shortsection{Secure Aggregation for Federated Learning}

Consider training a deep neural network to predict the next word that a
user will type as she composes a text message to improve typing
accuracy for a phone's on-screen keyboard \cite{goodman2002language}.
A modeler may wish to train such a model on all text messages across a
large population of users.
However, text messages frequently contain
sensitive information; users may be reluctant to upload a copy of them to
the modeler's servers.  Instead, we consider training
such a model in a \emph{Federated Learning} setting, wherein each user
maintains a private database of her text messages securely on her own
mobile device, and a shared global model is trained under the coordination
of a central server based upon highly processed, minimally scoped, ephemeral
updates from users \citep{mcmahan2016communication, shokri2015privacy}.

A neural network represents a function $f(x, \Parameters) = y$ mapping
an input $x$ to an output $y$, where $f$ is parameterized by a
high-dimensional vector $\Parameters \in \reals^\numElements$.  For
modeling text message composition, $x$ might encode the words entered
so far and $y$ a probability distribution over the next word.  A
training example is an observed pair $\langle x, y \rangle$ and a
training set is a collection $\Database = \left\{ \langle x_i, y_i
  \rangle; i = 1, \ldots, m \right\}$.  We define a loss on a training
set $\Loss_f(\Database, \Parameters) = \frac{1}{|\Database|}
\sum_{\langle x_i, y_i \rangle \in \Database} \Loss_f(x_i,
y_i, \Parameters)$, where $\Loss_f(x, y, \Parameters) = \norm(y,
f(x, \Parameters))$ for a loss function $\norm$, e.g., $\norm(y,
\hat{y}) = (y - \hat{y})^2$.
Training consists of finding parameters $\Parameters$
that achieve small $\Loss_f(\Database, \Parameters)$, typically
using a variant minibatch stochastic gradient descent \citep{chen16revisiting,goodfellow16deeplearning}.

In the Federated Learning setting, each user
$\user \in \Users$ holds a private set $\Database_\user$ of training
examples with $\Database = \bigcup_{\user \in \Users}
\Database_\user$. To run stochastic gradient descent, for each update
we select data from a random subset $\Users' \subset \Users$ and form
a (virtual) minibatch $\Batch = \bigcup_{\user \in \Users'}
\Database_\user$ (in practice we might have say $|\Users'| = 10^4$
while $|\Users| = 10^7$; we might only consider a subset of each
user's local dataset).
The minibatch loss gradient $\nabla \Loss_f(\Batch, \Parameters)$ can
be rewritten as a weighted average across users: $\nabla
\Loss_f(\Batch, \Parameters) = \frac{1}{|B|} \sum_{\user \in \Users'}
\Update_\user^t$ where $\Update_\user^t = |\Database_\user| \nabla
\Loss_f(\Database_\user, \Parameters^t)$.  A user can thus share just
$\langle |\Database_\user|, \Update_\user^t \rangle$ with the server,
from which a gradient descent step $\Parameters^{t+1}
\leftarrow \Parameters^t - \eta \frac{\sum_{\user \in \Users'}
  \Update_\user^t}{\sum_{\user \in \Users'} |\Database_\user|}$ may be
taken.
Although each update
$\langle |\Database_\user|, \Update_\user^t \rangle$ is ephemeral and contains less information
then the raw $\Database_\user$, a user might still
wonder what information remains. There
is evidence that a trained neural network's parameters sometimes allow reconstruction of
training examples~\citep{fredrikson2015model, shokri2015privacy, abadi2016deep};
might the parameter updates be subject to similar attacks?
For example, if the input $x$ is a one-hot vocabulary-length vector encoding
the most recently typed word, common neural network architectures will
contain at least one parameter $\theta_w$ in $\Parameters$ for each word $w$
such that $\frac{\partial \Loss_f}{\partial \theta_w}$ is non-zero only when $x$ encodes $w$.
Thus, the set of recently typed words in $\Database_\user$
would be revealed by inspecting the non-zero entries of $\Update_\user^t$.
The server does not need to inspect any individual user's update, however; it requires only the sums
$\sum_{\user \in \Users} |\Database_\user|$ and $\sum_{\user \in \Users} \Update_\user^t$.
Using a Secure Aggregation protocol would ensure that the server learns only that
\emph{one or more} users in $\Users$ wrote the word $w$,
but not \emph{which} users.

Federated Learning systems face several
practical challenges.
Mobile devices have only sporadic access to power and
network connectivity, so the set $\Users$ participating in each update step is unpredictable and the system must be robust to users dropping out.
Because $\Theta$ may contain millions of parameters,
updates $\Update_\user^t$ may be large, representing a direct cost to users
on metered network plans.  Mobile devices also generally cannot establish direct communications channels with other mobile devices (relying
on a server or service provider to mediate such communication) nor can they
natively authenticate other mobile devices.
Thus, Federated Learning motivates a need for a Secure Aggregation protocol that:
(1) operates on high-dimensional vectors,
(2) is communication efficient, even with a novel set of users on each instantiation,
(3) is robust to users dropping out, and
(4) provides the strongest possible security under the constraints
of a server-mediated, unauthenticated network model.

\shortsection{A Practical Secure Aggregation Protocol}

In our protocol, there are two
kinds of parties: a single server $\Server$ and a collection
of $\numUsers$ users $\Users$.  Each user $\user \in \Users$ holds a private vector $x_\user$ of dimension $\numElements$.
We assume that all elements of both $x_\user$ and $\sum_{\user \in \Users} x_\user$ are integers on
the range $[0, \maxElement)$ for some known $\maxElement$\footnotemark.
\footnotetext{Federated Learning updates $\Update_\user \in \reals^\numElements$ can be mapped
to $[0,\maxElement)^\numElements$ through a combination of clipping/scaling, linear transform, and (stochastic) quantization.}
Correctness requires that if all parties are honest, $\Server$ learns
$\bar{x} = \sum_{\user \in \bar{\Users}} x_\user$ for some subset of users
$\bar{\Users} \subseteq \Users$ where $|\bar{\Users}| \geq \frac{\numUsers}{2}$.
Security requires that
(1) $\Server$ learns nothing other than what is inferable from
$\bar{x}$, and (2) each user $\user \in \Users$ learns nothing.  We consider three different threat models. In all of them, all users follow the protocol honestly, but the server may attempt to learn extra information in different ways\footnote{We do not analyze security against arbitrarily malicious servers and users that may collude. We defer this case and a more formal security analysis to the full version.}:
\vspace{-8pt}
\begin{description} \itemsep-3pt
\item[(\threatmodel{1})]
The server is honest-but-curious, that is it follows the protocol honestly, but tries to learn as much as possible from messages it receives from users.

\item[(\threatmodel{2})]
The server can lie to users about which other users have dropped out, including reporting dropouts inconsistently among different users.

\item[(\threatmodel{3})]
The server can lie about who dropped out  (as in \threatmodel{2}) and also access the private memory of some limited number of users (who are following the protocol honestly themselves). (In this, the privacy requirement applies only to the inputs of the remaining users.)
\end{description}
\vspace{-8pt}

\shortparagraph{Protocol 0: Masking with One-Time Pads}
We develop our protocol in a series of refinements.  We begin by assuming that all parties complete the protocol and
possess pair-wise secure communication channels with ample bandwidth.
Each pair of
users first agree on a matched pair of input perturbations.
That is, user $\useri$ samples a vector $s_{\useri,\userj}$ uniformly from
$[0, \maxElement)^\numElements$ for each other user $\userj$.
Users $\useri$ and $\userj$ exchange $s_{\useri,\userj}$ and $s_{\userj,\useri}$ over their secure
channel and compute perturbations $p_{\useri,\userj} = s_{\useri,\userj} - s_{\userj,\useri} \pmod{\maxElement}$,
noting that $p_{\useri,\userj} = -p_{\userj,\useri} \pmod{\maxElement}$
and taking $p_{\useri,\userj}=0$ when $\useri = \userj$.
Each user sends to the server:
$y_{\useri} = x_{\useri} + \sum_{\userj \in \Users} p_{\useri,\userj} \pmod{\maxElement}$.
The server simply sums the perturbed values:
$\bar{x} = \sum_{\user \in \Users} y_{\user} \pmod {\maxElement}$.  Correctness is guaranteed
because the paired perturbations in $y_{\user}$ cancel:
\begin{equation*}
\bar{x} =
\sum_{\user \in \Users} x_{\user} + \sum_{\useri \in \Users} \sum_{\userj \in \Users} p_{\useri,\userj} =
\sum_{\user \in \Users} x_{\user} + \sum_{\useri \in \Users} \sum_{\userj \in \Users} s_{\useri,\userj} - \sum_{\useri \in \Users} \sum_{\userj \in \Users} s_{\userj,\useri} =
\sum_{\user \in \Users} x_{\user} \pmod{\maxElement}.
\end{equation*}

Protocol 0 guarantees perfect privacy for the users;
because the $s_{\useri,\userj}$ factors that users add are uniformly sampled, the $y_\useri$ values appear uniformly random to the server, subject to the constraint that $\bar{x} = \sum_{\user \in \Users} y_{\user} \pmod {\maxElement}$. In fact, even if the server can access the memory of some users, privacy holds for those remaining.
\footnote{A more complete and formal argument is deferred to the full version of this paper.}

\shortparagraph{Protocol 1: Dropped User Recovery using Secret Sharing}

Unfortunately, Protocol 0 fails several of our design criteria,
including robustness:
if any user $\useri$ fails to complete the protocol by sending her $y_{\useri}$
to the server, %
the resulting
sum will be masked by the perturbations that $y_\useri$ would have cancelled.
To achieve robustness, we first add an initial round to the protocol in which
user $\useri$ generates a public/private keypair,
and broadcasts the public key over the pairwise channels.
All future messages from $\useri$ to $\userj$
will be intermediated by the server but encrypted with $\userj$'s public key, and signed by $\useri$, simulating a secure authenticated channel.
This allows the server to maintain a consistent view of which users have
successfully passed each round of the protocol. (We assume here, temporarily, that the server faithfully delivers all messages between users.)

We also add a secret-sharing round between users after
$s_{\useri,\userj}$ values have been selected.  In this round, each
user computes $\numUsers$ shares of each perturbation $p_{\useri,\userj}$ using a
$(\threshold, \numUsers)$-threshold scheme
\footnote{A $(\threshold, \numUsers)$ secret-sharing scheme allows splitting a secret into $\numUsers$ shares, such that any subset of $\threshold$ shares is sufficient to recover the secret, but given any subset of fewer than $\threshold$ shares the secret remains completely hidden.}, such as Shamir's Secret
Sharing~\citep{shamir1979share}, for some $\threshold > \frac{\numUsers}{2}$.
For each secret
user $\useri$ holds, she encrypts one share with each user $\userj$'s public key, then delivers all of these shares to the server.
The server gathers
shares from a subset of the users $\Users_1 \subseteq \Users$ of size at least $t$ (e.g. by waiting a for a fixed period),
then considers all other users dropped. The server delivers to each user $\userj \in \Users_1$
 the secret shares that were encrypted for that user; all
the users in $\Users_1$ now infer a consistent view of the surviving user set $\Users_1$ from the set of received shares.  When a user computes $y_{\user}$, she only includes those perturbations related
to surviving users; that is,
$y_{\useri} = x_{\useri} + \sum_{\userj \in \Users_1} p_{\useri,\userj} \pmod{\maxElement}$.

After the server has received $y_{\user}$ from at least $\threshold$ users $\Users_2 \subseteq \Users_1$, it
proceeds to a new unmasking round, considering all other users to be dropped.
From the remaining users in $\Users_2$, the
server requests all shares of secrets generated by the dropped users in $\Users_1 \setminus \Users_2$.
As long as $|\Users_2| > t$, each user will respond with those shares.  Once the server receives shares from at least $\threshold$ users, it
reconstructs the perturbations for $\Users_1 \setminus \Users_2$ and
computes the aggregate value:
$\bar{x} = \sum_{\user \in \Users_2} y_{\user} - \sum_{\useri \in \Users_2} \sum_{\userj \in \Users_1 \setminus \Users_2} p_{\useri,\userj}\pmod {\maxElement}$.
Correctness is guaranteed for $\bar{\Users} = \Users_2$ as long as at least $\threshold$ users complete the protocol. In this case, the sum $\bar{x}$ includes the values of at least $\threshold > \frac{\numUsers}{2} $ users, and all perturbations cancel out:
\begin{equation*}
\bar{x}
= \left(\sum_{\user \in \Users_2} x_{\user} + \sum_{\useri \in \Users_2} \sum_{\userj \in \Users_1} p_{\useri,\userj}\right)
  - \sum_{\useri \in \Users_2} \sum_{\userj \in \Users_1 \setminus \Users_2} p_{\useri,\userj}
= \sum_{\user \in \Users_2} x_{\user} + \sum_{\useri \in \Users_2} \sum_{\userj \in \Users_2} p_{\useri,\userj}
= \sum_{\user \in \Users_2} x_{\user} \pmod{\maxElement}.
\end{equation*}
However, security has been lost: if a server incorrectly omits $\useri$ from $\Users_2$,
either inadvertently (e.g. $y_{\useri}$ arrives slightly too late) or by
malicious intent, the honest users in $\Users_2$ will supply
the server with all the secret shares needed to remove all the perturbations that
masked $x_{\useri}$ in $y_{\useri}$. This means we cannot guarantee security even against honest-but-curious servers  (Threat Model \threatmodel{1}).

\shortparagraph{Protocol 2: Double-Masking to Thwart a Malicious Server}
To guarantee security, we introduce a double-masking
structure that protects $x_{\useri}$ even when the server can reconstruct $\useri$'s perturbations.
First, each user $\useri$ samples an additional
random value $b_{\useri}$ uniformly from $[0, \maxElement)^\numElements$ during the same round as the
generation of the $s_{\useri,\userj}$ values.  During the secret sharing round,
the user also generates and distributes shares of $b_{\useri}$ to each of
the other users.  When generating
$y_{\useri}$, users also add this secondary mask: $y_{\useri} = x_{\useri} + b_{\useri} + \sum_{\userj \in \Users_1} p_{\useri,\userj} \pmod{\maxElement}$.
During the unmasking round, the server must make an explicit choice with respect to each user $\useri \in \Users_1$: from each surviving member $\userj \in \Users_2$, the server can request  \emph{either}
a share of the $p_{\useri,\userj}$ perturbations associated with $\useri$ \emph{or}
a share of the $b_{\useri}$ for $\useri$; an honest user $\userj$ will only respond if $|\Users_2| > t$, and will never reveal both kinds of shares
for the same user.  After gathering at least $\threshold$ shares of $p_{\useri,\userj}$ for all $\useri \in \Users_1 \setminus \Users_2$ and $\threshold$ shares of $b_\user$ for all $\user \in \Users_2$, the server
reconstructs the secrets and computes the aggregate value:
$\bar{x} = \sum_{\user \in \Users_2} y_{\user} - \sum_{\user \in \Users_2} b_{\useri} - \sum_{\useri \in \Users_2} \sum_{\userj \in \Users_1 \setminus \Users_2} p_{\useri,\userj}\pmod {\maxElement}$.

We can now guarantee security in Threat Model \threatmodel{1} for $\threshold > \frac{\numUsers}{2}$, since $x_\useri$ always remains masked by either $p_{\useri, \userj}$s or by $b_\useri$s.
It can be shown that in Threat Models \threatmodel{2} and \threatmodel{3} the thresholds must be raised to $\frac{2\numUsers}{3}$ and $\frac{4\numUsers}{5}$ correspondingly.
We defer the detailed analysis, as well as the case of arbitrarily malicious and colluding servers and users, to the full version\footnotemark.
\footnotetext{The security argument involves bounding the number of shares the server can recover by forging dropouts.}

\shortparagraph{Protocol 3: Exchanging Secrets Efficiently}

While Protocol 2 is robust and secure with the right choice of $\threshold$, it requires $O(\numElements\numUsers^2)$ communication, which
we address in this refinement of the protocol.
Observe that a single secret value may be expanded to a vector of pseudorandom values by using it to seed a cryptographically secure pseudorandom generator
(PRG)~\citep{acs2011dream, golle2004dining}.  Thus we can generate just scalar seeds $s_{\useri, \userj}$ and
$b_\useri$ and expand them to $\numElements$-element vectors.
Still, each user has $(\numUsers-1)$ secrets
$s_{\useri, \userj}$ with other users and must publish shares of all these secrets.  We use key agreement to establish these
secrets more efficiently.  Each user generates a Diffie-Hellman secret key $s^{SK}$ and
public key $s^{PK}$.  Users send their public keys to the server
(authenticated as per Protocol 1); the server
then broadcasts all public keys to all users, retaining a copy for itself.
Each pair of users $\useri, \userj$ can now agree on a secret
$s_{\useri, \userj} = s_{\userj, \useri} = \Agree(s^{SK}_\useri, s^{PK}_\userj) = \Agree(s^{SK}_\userj, s^{PK}_\useri)$.
To construct perturbations, we assume a total ordering
on $\Users$ and take $p_{\useri, \userj} = \PRG(s_{\useri, \userj})$ for
$\useri < \userj$, $p_{\useri, \userj} = -\PRG(s_{\useri, \userj})$ for
$\useri > \userj$, and $p_{\useri, \userj}=0$ for $\useri = \userj$ (as before).
The server now only needs to learn $s^{SK}_\user$ to reconstruct all of
$\user$'s perturbations; therefore $\user$ need only distribute shares of
$s^{SK}_\user$ and $b_\user$ during the secret sharing round.
The security of Protocol 3 can be shown to be essentially identical to that of Protocol 2 in each of the different threat models.

\shortparagraph{Protocol 4: Minimizing Trust in Practice}
\begin{figure}[t]
\begin{minipage}{.35\textwidth}
\setlength{\belowcaptionskip}{-12pt}
  \centering
  \begin{tabular}{ll}
  \hline
  \multicolumn{2}{l}{\textbf{computation}} \\
  User & $O(\numUsers^2 + \numElements\numUsers)$ \\
  Server\footnotemark & $O(\numElements\numUsers^2)$\\
  \multicolumn{2}{l}{\textbf{communication}} \\
  User & $O(\numUsers + \numElements)$ \\
  Server & $O(\numUsers^2 + \numElements\numUsers)$ \\
  \multicolumn{2}{l}{\textbf{storage}} \\
  User & $O(\numUsers + \numElements)$ \\
  Server & $O(\numUsers^2 + \numElements)$ \\
  \hline
  \end{tabular}
  \captionof{table}{%
   Protocol 4 Cost Summary (derivations deferred to the full paper).
 }\label{table:costs}
\end{minipage}
\begin{minipage}{.56\textwidth}
\setlength{\belowcaptionskip}{-12pt}
\includegraphics[width=\textwidth]{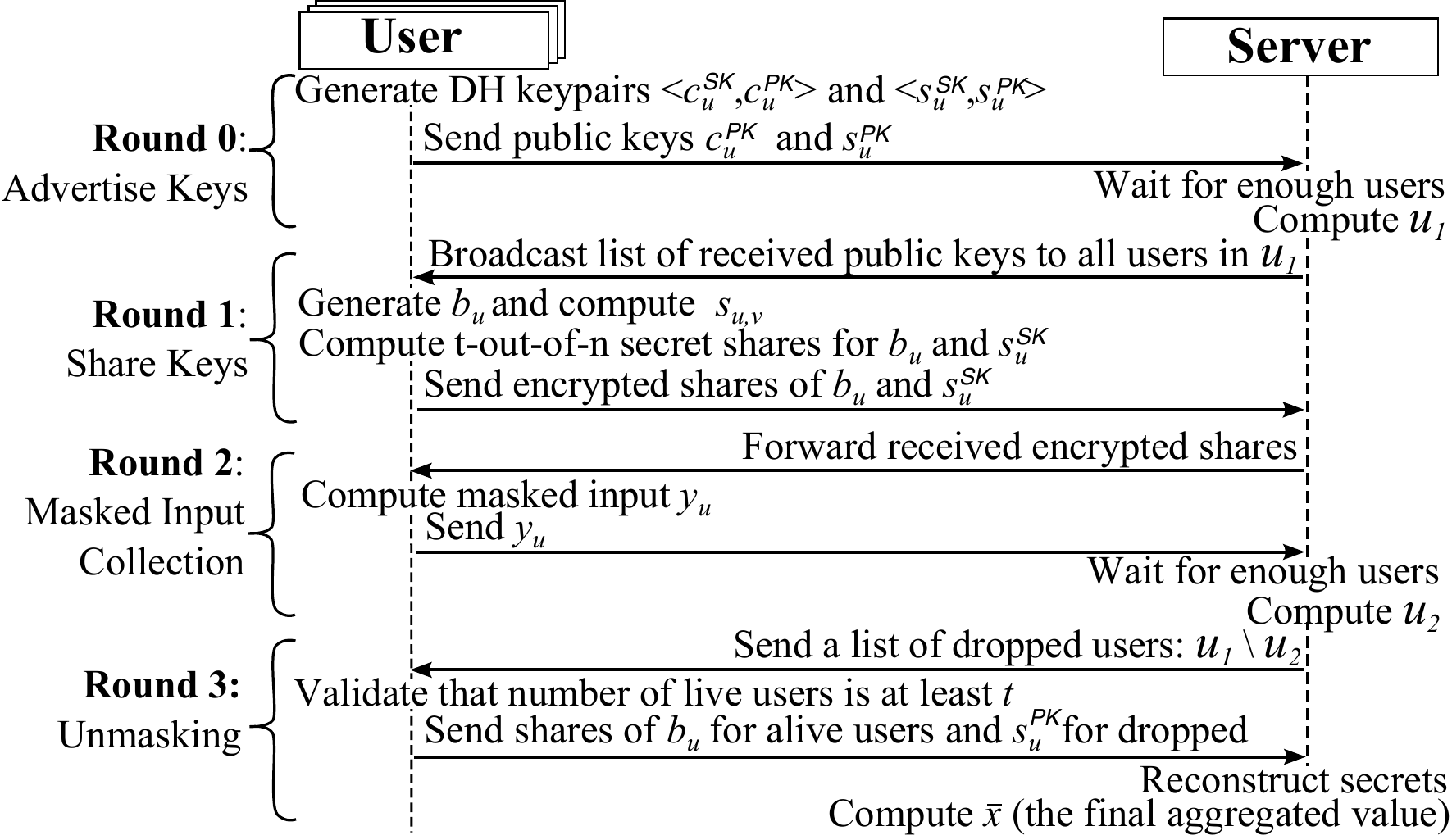}
\captionof{figure}{Protocol 4 Communication Diagram}
\end{minipage}
\end{figure}
\footnotetext{We reconstruct $\numUsers$ secrets from aligned $(\threshold,\numUsers)$-Shamir shares in $O(\threshold^2 + \numUsers\threshold)$ by caching Lagrange coefficients.}

Protocol 3 is not practically deployable for mobile devices because they
lack pairwise secure communication and authentication.
We propose to bootstrap the communication protocol by replacing
the exchange of public/private keys described in Protocol 1 with a server-mediated
key agreement, where each user generates a Diffie-Hellman secret key $c^{SK}$ and
public key $c^{PK}$ and advertises the latter together with $s^{PK}$\footnote{This can be viewed as bootstrapping a SSL/TLS connection between each pair of users}.
We note immediately that the server may now conduct man-in-the-middle attacks,
but argue that this is tolerable for several reasons.
First, it is essentially inevitable for users that lack
authentication mechanisms or a pre-existing public-key infrastructure.  Relying only on the non-maliciousness of the
bootstrapping round also constitutes minimization of trust: the code
implementing this stage is small and could be publicly audited,
outsourced to a trusted third party, or implemented via a
trusted compute platform offering a remote attestation
capability~\citep{costansanctum,costan2016sgx,suh2003aegis}. Moreover,
the protocol meaningfully increases security
(by protecting against anything less than an actively malicious attack
by the server) and provides forward secrecy (compromising the server at any
time after the key exchange provides no benefit to the attacker, even if
all data and communications had been fully logged).

We summarize the protocol's performance in Table~\ref{table:costs}.
Taking that key agreement public keys and encrypted secret shares
are 256 bits and that users' inputs are all on the same
range\footnote{Taking $\maxElement=\numUsers(\maxElement_U-1)+1$ to
  ensure no overflow} $[0, \maxElement_U-1]$, each user transfers
$\frac{256 (7\numUsers-4) + \numElements \ceil*{\log_2
    \left(\numUsers(\maxElement_U-1)+1\right)} +
  \numUsers}{\numElements \ceil*{\log_2 \maxElement_U}}$ more data
than if she sent a raw vector.

\label{Section.Evaluation}

\shortsection{Related work}
The restricted case of secure aggregation in which all users but one have an input 0 can be expressed as a dining cryptographers network (DC-net), which provide anonymity by using pairwise blinding of inputs~\citep{chaum1988dining, golle2004dining}, allowing to untraceably learn each user's input.
Recent research has examined the communication efficiencly and operation in the presence of malicious users~\citep{corrigangibbs2013proactively}. However, if even one user aborts too early, existing protocols must restart from scratch, which can be very expensive~\citep{kwon2015riffle}.
Pairwise blinding in a modulo addition-based encryption scheme has been explored, but existing schemes are neither efficient for vectors nor robust to even single failure~\citep{acs2011dream, goryczka2015comprehensive}.  Other schemes (e.g. based on Paillier cryptosystem~\citep{rastogi2010differentially}) are very computationally expensive.

\bibliographystyle{plainnat}
\bibliography{refs}

\end{document}